\documentstyle[12pt]{article}
\def\setb@se#1{\baselineskip=#1 \normalbaselineskip=#1}
\setb@se{14pt}
\newcommand{\be}{\begin{equation}}
\newcommand{\ee}{\end{equation}}
\newcommand{\ii}{{\it i}}
\newcommand{\f}{\varphi}
\newcommand{\iii}{\int_{0}^{\infty}}

\begin{document}

\input epsf

\begin{titlepage}
\begin{flushright}
ZU-TH 12/95

hep-th/9609201
\end{flushright}

\vspace{40 mm}
\begin{center}
{\Huge \bf Negative modes around\\

Einstein--Yang--Mills sphalerons and black holes\footnote{
Published in {\em Geometry and Integrable Models}, Eds. 
P.N. Pyatov \& S.N.Solodukhin, World Scientific 1996, 
pp. 55-77.}

}

\vspace{10 mm}
\end{center}
\begin{center}
{\bf  Mikhail S. Volkov}
\vskip2mm
{\em Institut    f\"ur    Theoretische    Physik    der    Universit\"at
Z\"urich--Irchel, \\
Winterthurerstrasse 190, CH--8057 Z\"urich,
Switzerland,\\
e--mail: volkov@physik.unizh.ch}
\vskip1cm
\end{center}

\vspace{20mm}
\begin{center}{\bf Abstract}\end{center}
\vspace{2mm}

The dynamics of small perturbations around 
sphaleron and black hole
solutions in the Einstein-Yang-Mills theory 
for the
gauge group $SU(2)$ is investigated.
The perturbations can be split into the two
independent sectors in accordance with their parity;
each sector contains negative modes.
The even-parity negative modes are shown to
correspond to the negative second variations of the
height of the potential energy barrier near the
sphaleron.  For the odd-parity sector,
the existence of precisely $n$
(the number of nodes of the sphaleron solution gauge
field function) 
negative  modes is rigorously proven.
The same results hold for the Einstein-Yang-Mills
black holes as well.

\end{titlepage}

\section{Introduction}

Soon after the discovery of the Bartnik-McKinnon solitons \cite{BK}
and black holes \cite{BH}  in the $SU(2)$ Einstein-Yang-Mills
(EYM) theory, it was shown that these interesting objects are
unfortunately unstable \cite{SZ}. However, as has been realized later
\cite{sphal}, the property to be unstable is in fact fairly
important in the case, since it allows for the sphaleron
interpretation \cite{manton},\cite{MK} of the regular BK objects.

Usually, sphalerons are related with the existence of
non-contractible loops in the configuration space of the theory
\cite{manton}, \cite{forgacs}. For each such a loop, one finds a
maximal value of the potential energy and then minimizes
the result over all loops; this gives the energy of a static
saddle point field configuration. Notice that the
configuration space has to be compact in this approach,
otherwise the existence of non-contractible loops does not
ensure the existence of saddle points of the energy \cite{manton}.
The difference between the two cases is perfectly illustrated
in the two figures in famous Manton's paper
(see Fig.1 and Fig.2 in \cite{manton}). Although for the
non-compact case the saddle points may also exist
(see Fig.2 in \cite{manton}), these points have {\it more then
one} directions of instability, and thus can not be constructed
in the same manner is in the compact case, such that their
existence is not directly related to the existence of
non-contractible loops. The use of loops in the non-compact case
may have another meaning: for theories with vacuum periodicity,
the non-contractible loop through the saddle point (if any exists)
can be transformed into a path interpolating between distinct vacua.
This shows that the saddle point field configuration
relates to the top of the potential barrier between the vacua
thus justifying the usage of the name ``sphaleron''.    

It is not widely understood that the EYM sphalerons correspond to
the ``non-compact'' case.
In the EYM theory, the maximal value of energy for each
non-contractible loop can be reduced to zero by a smooth
deformation of the loop \cite{sphal}, such that the existence
of the loops {\it does not ensure} the existence of the solutions
(otherwise the existence proof for the BK solutions \cite{SM} --
\cite{BFM} would just follow the same line as
that for the standard sphaleron \cite{burz}).
Instead, in order to illustrate the sphaleron nature
of the solutions, 
the loops can be used as paths connecting the
solutions with the neighboring vacua \cite{level} (see
below).  Manifestly, the difference between the
standard Yang-Mills-Higgs (YMH) sphaleron and the EYM sphalerons
results in the different numbers of their unstable modes.
The YMH sphaleron \cite{manton} has one and only one negative mode
in the odd-parity spherically-symmetric perturbation sector.
In the even-parity sector, the solution has no negative modes,
such that the energy of the solution indeed defines
the {\it minimal} height of the barrier.
The EYM sphalerons have negative modes both in the odd parity
and in the even-parity sectors, which means that they
do not specify the minimal possible height of the barrier
(the latter is zero) \cite{small}.

The present paper concerns with the investigation of the
dynamics of
spherically-symmetric perturbation modes around
EYM sphalerons and black holes. The perturbations decouple
into the two independent groups in accordance with their parity.
The even parity modes were studied previously in
\cite{SZ}, it has been found numerically that the
$n$-th soliton or black hole solution has precisely $n$
negative modes of this type. Here we explicitly show
that the negative modes of this type are related to
the decrease of the {\it height} of the potential
barrier around the sphaleron \cite{small}.
For the odd-parity modes,
the existence of at least one such a mode for all
known $SU(2)$ EYM solitons and black holes has been established
in \cite{sphal}, \cite{odd}, as well as for the "generic''
solitons in the EYM theory for any compact gauge group \cite{BS}.
For the lowest $n$ regular BK solutions, 
numerical analysis has revealed the existence of $n$ negative modes
of this type too \cite{LM}.
Here  we give a complete proof of the following statement:
for all known regular and black hole non-Abelian EYM solutions
there exist precisely $n$ negative modes in the odd parity
spherically-symmetric perturbation sector \cite{super}.
For the even-parity modes the corresponding proof is
unknown, however the results of the numerical analysis
suggest that there are $n$ negative modes in this case too,
such that the total number of negative modes for
the EYM solitons and black holes is $2n$.

The rest of the paper is organized as follows.
In Sec.2 the basic notations and equations are introduced.
Sec.3 contains an explicit construction of vacuum-to-vacuum
paths through the EYM sphalerons. In Sec.4 the basic
perturbation equations are given for the both parity
types. For the odd-parity perturbation modes, the
transformation to a one-channel Schr\"odinger problem
is given in Sec.5, the regularity
of the effective potential is shown in Sec.6.
A zero-energy bound state
in the problem  is explicitly presented in Sec.7,
the corresponding wave function has $n$ zeros, which
proves the existence of precisely $n$ bound states.
The black hole case is considered in Sec.8,
the results apply also for the case of charged
non-Abelian black holes found in the EYM theory
with the $SU(2)\times U(1)$ gauge group \cite{charge}.
The second variation of the ADM mass functional
for the regular BK solutions is explicitly derived in
the Appendix.

\section{EYM field equations}

We start from the action of the $SU(2)$ EYM theory 
\be
S=-{1\over 16\pi G}\int R\sqrt{-g}d^{4}x - 
{1\over 2e^{2}}\int trF_{\mu\nu}F^{\mu\nu}
\sqrt{-g}d^{4}x,                     \label{1}
\ee
where       $F_{\mu\nu}=\partial_{\mu}A_{\nu}-\partial_{\nu}A_{\mu}- 
\ii [A_{\mu},A_{\nu}]$ is the matrix valued gauge   field    tensor, 
$e$       is       the       gauge         coupling        constant, 
$A_{\mu}=A_{\mu}^{a}\tau^{a}/2$, and  $\tau^{a}\   (a=1,2,3)$    are 
the  Pauli  matrices.

In the spherically  symmetric   case, it is convenient to represent  
the   spacetime   metric in the following form:
\be
ds^{2} = l^{2}_{e}\left( (1-\frac{2m}{r})\sigma^{2}dt^{2} -
\frac{dr^{2}}{1-2m/r} - 
r^{2}(d\vartheta^{2} + \sin^{2}\vartheta
d\varphi^{2})\right),                                     \label{2}
\ee
where $m$ and $\sigma$ depend  on  $t$  and   $r$.  
Here the EYM length 
scale $l_{e}=\sqrt{4\pi G}/e$ is explicitly displayed, after such
a rescaling, 
all other quantities in the theory become dimensionless.

The  spherically  symmetric 
$SU(2)$ Yang-Mills field can be parameterized by
\be
A = W_{0}\hat{L}_{1}\ dt + W_{1}\hat{L}_{1}\ dr
+\{p_{2}\ \hat{L}_{2} - (1-p_{1})\
\hat{L}_{3}\}\ d\vartheta
+\{(1-p_{1})\ \hat{L}_{2} + p_{2}\
\hat{L}_{3}\}\sin\vartheta\ d\varphi,
                                                        \label{3}
\ee
where  $W_{0}$, $W_{1}$, $p_{1}$, $p_{2}$ are functions of  $t$  and 
$r$,           $\hat{L}_{1}           =            n^{a}\tau^{a}/2$, 
$\hat{L}_{2}=\partial_{\vartheta}\hat{L}_{1}$, 
$\sin\vartheta\hat{L}_{3}=\partial_{\varphi}\hat{L}_{1}$,        and 
$n^{a}=(\sin\vartheta          \cos\varphi,\sin\vartheta\sin\varphi, 
\cos\vartheta)$.  The  gauge  transformation 
\be
A \rightarrow\  UAU^{-1}+ \ii UdU^{-1}, \ \ {\rm with}\ \
U=\exp(\ii\Omega(t,r)\hat{L}_{1}),                        \label{4}
\ee
preserves the form of the field (\ref{3}),  altering  the  functions 
$W_{0}$, $W_{1}$, $p_{1}$, $p_{2}$ as 
\be
W_{0}\rightarrow W_{0}+\partial_{t}\Omega,\ \ \
W_{1}\rightarrow W_{1}+\partial_{r}\Omega,\ \ \
p_{\pm}=p_{1}\pm\ii   p_{2}
 \rightarrow \exp(\pm\ii\Omega)p_{\pm}.                    \label{5}
\ee
Consider also parity transformation: $\vartheta\rightarrow\pi -
\vartheta$, $\varphi\rightarrow\varphi +\pi$.
The action of this transformation on the fields (\ref{2}), (\ref{3})
is equivalent to the following replacement:
\be
m\rightarrow m,\ \
\sigma\rightarrow\sigma,\ \
p_{1}\rightarrow p_{1},\ \
p_{2}\rightarrow -p_{2},\ \
W_{0}\rightarrow -W_{0},\ \
W_{1}\rightarrow -W_{1}.                               \label{5:1}
\ee

It is convenient to  introduce  the  complex  variable  $f=p_{1}+\ii 
p_{2}$ and its covariant derivative $D_{\mu}f =  (\partial_{\mu}-\ii 
W_{\mu})f$,   as   well    as    the   $(1+1)$-dimensional     field 
strength  $W_{\mu\nu}= 
\partial_{\mu}W_{\nu}-\partial_{\nu}W_{\mu}$,  $(\mu,   \nu  =0,1)$. 
The full system of the EYM  equations then reads 
\be
\partial_{\mu} (r^{2}\sigma W^{\mu\nu})-2\sigma~ Im~ 
(fD^{\nu}f)^{\ast}=0,                                    \label{6}
\ee
\be
D_{\mu}\sigma D^{\mu}f-\frac{\sigma}{r^{2}}(|f|^{2}-1)f=0, 
                                                           \label{7}
\ee
\be
\partial_{r} m=-\frac{r^{2}}{4}W_{\mu\nu}W^{\mu\nu} +
\frac{1}{N\sigma^{2}}|D_{0}f|^{2}+N|D_{1}f|^{2}+
\frac{1}{2r^{2}}(|f|^{2}-1)^{2},                           \label{8}
\ee
\be
\partial_{t}m=2N~Re~D_{0}f(D_{1}f)^{\ast},                 \label{9}
\ee
\be
\partial_{r}\ln\sigma
= \frac{2}{r} \left(\frac{1}{N^{2}\sigma^{2}}
|D_{0}f|^{2}+|D_{1}f|^{2}\right),                         \label{10}
\ee
where $N=1-2m/r$, asterisk  denotes complex conjugation. 

In the static case, with $W_{0}=W_{1}= p_{2}=0$ and  $f=Re~f=p_{1}$, 
these  equations  are  known  to  possess  regular,
asymptotically flat solutions discovered by Bartnik and McKinnon (BK)
\cite{BK}, as well as
the EYM  black  hole solutions \cite{BH}.
The regular solutions form a discrete family  labeled  by 
an integer  $n=1,2,\ldots$.  
Near the origin and infinity, these solutions,
$(f_{n}(r),  m_{n}(r) \sigma_{n}(r))$, 
have the following common behaviour:
\be
f=1-br^{2}+O(r^{4}),\ \ m=2b^{2}r^{3}+O(r^{5}),\ \ 
\frac{\sigma}{\sigma_{0}}=1+4b^{2}r^{2}+O(r^{4}),
\ \ \ {\rm as}\ \ \
r\rightarrow 0,                               \label{a1}
\ee
and
\be
f=(-1)^{n}(1-\frac{a}{r}+O(\frac{1}{r^{2}})),\ \ 
m=M+O(\frac{1}{r^{3}}),\ \ \
\sigma=1+O(\frac{1}{r^{4}})\ \ 
{\rm as}\ \ r\rightarrow\infty,                     \label{a2}
\ee
where  $0<b,  \sigma_{0}<1$,  $M$  and  $a$  are  numerically   known 
constants depending on $n$. In  the  domain   $0<r<\infty$,   $m(r)$  
and  $\sigma(r)$   are   monotone   increasing   functions,    while   
$f(r)$ oscillates $n$  times  around  zero  value   always   staying  
within  the stripe $-1<f(r)<1$.

The EYM black   hole   solutions   are   distinguished   by  the two 
parameters,  $n$ and $r_{h}$, where $n$ is the number of  nodes   of  
the   function    $f(r)$    in    the    region    $r>r_{h}$,    and 
$r_{h}\in(0,\infty)$  is  the  event  horizon   size.  The  boundary 
conditions at the horizon $r=r_{h}$ are 
\be
f=f_{h}+\frac{P}{F}~x+O(x^{2}),\ \
N=F~x+O(x^{2}),\ \ 
\frac{\sigma}{\sigma_{h}}= 1+
\frac{2P^{2}}{r_{h}^{2}F^{2}}~x+O(x^{2}),            \label{a3}
\ee
where
\be
x=\frac{r-r_{h}}{r_{h}},\ \ \
P=f_{h}(f_{h}^{2}-1),\ \ \
F=1-(f_{h}^{2}-1)^{2}/r_{h}^{2},\ \                     \label{a3:1}
\ee
and $0<f_{h}, \sigma_{h}$  depend  on  $(n,r_{h})$.  In  the 
domain  $r_{h}<r<\infty$  the  behaviour   of   these solutions   is 
qualitatively 
similar to that for the regular case with the boundary conditions at 
infinity defined by (\ref{a2}).

In what follows, it will be
convenient to  introduce   the   tortoise   coordinate  $\rho$  
such  that $d\rho/dr=1/\sigma N$. In the regular case  $\rho$   runs  
over the half-line $[0,\infty)$: 
\be
\rho=\frac{r}{\sigma_{0}}+O(r^{5})\ \ {\rm as}\ \ 
r\rightarrow 0,\ \ \ \ \ \ \ \ 
\rho=r+2M\ln r+O(1)\ \ 
{\rm as}\ \ r\rightarrow\infty.                        \label{a4}
\ee
For the black hole solutions the  domain  of  $\rho$  is  the  whole  
line $(-\infty,+\infty)$ such that near the horizon 
\be
N=1-\frac{2m}{r}=F\exp\left(\sigma_{h}F\frac{\rho}{r_{h}}\right)+
O\left(\exp\left(2\sigma_{h}F\frac{\rho}{r_{h}}\right)\right)\ \ 
{\rm as}\ \ \rho\rightarrow -\infty.                  \label{a5}
\ee

\section{EYM sphalerons and vacuum to vacuum paths}

As was shown in Ref.\cite{sphal},  the  odd-$n$  BK  solutions  may  be 
treated as sphalerons. This interpretation  is  based  on  the  fact 
that, through any such a solution, one may find a one-parameter family 
of static field configurations which  interpolates  between  topologically
distinct  EYM 
vacua. If  $\lambda$  denotes  the  parameter  of  the  family,  the 
corresponding gauge field can be chosen as follows \cite{sphal}
\be 
A_{\mu}dx^{\mu}=\frac{\ii}{2e}(1-f_{n}(r))U dU^{-1},\ \ \ 
U=exp(\ii\lambda n^{a}\tau^{a}),                          \label{l7}
\ee
where $\lambda$ ranges from zero to $\pi$. The metric functions  $m$ 
and $\sigma$ in (\ref{2}) are given by 
$$ 
\sigma (r)= exp\{-2sin^{2}\lambda\int_{r}^{\infty}f_{n}'^{2} 
\frac{dr}{r}\}, $$
\be
m(r)=\frac{sin^{2}\lambda}{\sigma(r)}\int_{0}^{r}(f_{n}'^{2} + 
sin^{2}\lambda\frac{(f_{n}^{2}-1)^{2}}{2r^{2}})\sigma dr. \label{l8} 
\ee
It is worth noting that, for any value of $\lambda$, the  fields  of 
this family  satisfy  the  two  Einstein  equations  $G^{0}_{0}=8\pi 
GT^{0}_{0}$ and $G^{r}_{r}=8\pi  GT^{r}_{r}$;  when  $\lambda=\pi/2$ 
these equations reduce to Eqs.(\ref{8}), (\ref{10}) \cite{sphal}.
One can check that, when $\lambda=\pi/2$, the fields defined by
(\ref{l7}), (\ref{l8})  coincide  with the $n$-th BK solution field
described above. 
When  $\lambda$  approaches  zero  or 
$\pi$ values, the spacetime metric becomes flat and the gauge  field 
(\ref{l7}) vanishes. Thus, the family of  fields  (\ref{l7}),
(\ref{l8}) 
describes a loop (in the EYM  configuration  function  space)  which 
interpolates between the trivial vacuum and the $n$-th  BK  solution 
when $\lambda$ runs from 0 to  $\pi/2$,  and  returns  back  to  the 
vacuum as $\lambda$ changes from $\pi/2$ to $\pi$. 

It is important  that  this  loop  can  be  transformed  to  a  path 
interpolating between neighboring  EYM vacua with different winding 
numbers of the gauge field. To see this, one has to pass to such a gauge,
in which the gauge field potential decays faster then $1/r$ as
$r\rightarrow\infty$. The gauge transformation
\be
A_{\mu}\rightarrow U(A_{\mu}+\frac{\ii}{e}\partial_{\mu})U^{-1},\ \ 
{\rm with}\ \ 
U=exp (\ii\frac{\lambda}{2}(f_{n}-1)n^{a}\tau^{a})
                                               \label{l8:1}
\ee
yields 
$$
A_{\mu}dx^{\mu}=\frac{\ii}{2e}(1-f_{n}(r))U_{+}\ dU_{+}^{-1}+
\frac{\ii}{2e}(1+f_{n}(r))U_{-}\ dU_{-}^{-1},
$$
\be
U_{\pm}=exp(\ii\lambda (f_{n}\pm 1)n^{a}\tau^{a}/2),     \label{l9}
\ee
whereas  the  gauge  invariant  metric  functions  (\ref{l8})  remain 
unchanged. One can see that, when $\lambda$ runs from zero to $\pi$, 
the field (\ref{l9}) interpolates between two different pure  gauges. 
Notice that, in the new gauge, one has  $A_{a}=O(r)$ 
when     $r\rightarrow     0$     and     $A_{a}=O(1/r^{2})$      as 
$r\rightarrow\infty$, with $A_{a}$ being the  field  component  with 
respect to an orthonormal frame. 

Let $\lambda$ depend adiabatically on  time,  in  such  a  way  that 
$\lambda(t= -\infty)=0$ and $\lambda(t=\infty)=\pi$. Notice 
that the structure of the field (\ref{l9}) ensures the
temporal gauge condition $A_{0}=0$ even for time-dependent $\lambda$.
Recall the Chern-Simons  (CS)  current and its divergence
\be
K^{\mu}={e^{2}\over 8\pi^2}tr{\varepsilon^{\mu\nu\alpha\beta}\over 
\sqrt{-g}}A_{\nu}(\nabla_{\alpha}A_{\beta}-\frac{2\ii e}{3}
A_{\alpha}A_{\beta}),\ \ \nabla_{\alpha}K^{\alpha}=
\frac{e^{2}}{16\pi^{2}}trF_{\mu\nu}\tilde{F}^{\mu\nu},   \label{l10}
\ee
where   $\nabla_{\alpha}$   is   the   covariant   derivative,   and 
$\tilde{F}^{\mu\nu}$  is  dual  tensor.  Integration  leads  to  the 
following relation:
\be
\frac{e^{2}}{16\pi^{2}}\int_{-\infty}^{t}dt\int d^{3}x \sqrt{-g}
trF_{\mu\nu}\tilde{F}^{\mu\nu}=
\left.\int d^{3}x\sqrt{-g}K^{0}\right|^{t}_{-\infty} 
+ \int_{-\infty}^{t} dt\oint\vec{K}d\vec{\Sigma}.        \label{l11}
\ee
In the temporal gauge used, the field decays faster then $1/r$ as
$r\rightarrow\infty$, such that the surface integral entering the
right
hand side vanishes, while the gauge  invariant  left  hand  side  is 
\cite{sphal}
\be
\frac{3}{2\pi}\int_{-\infty}^{t}dt\ \dot{\lambda}sin^{2}\lambda
\int_{0}^{\infty}dr\ f_{n}'(f_{n}^{2}-1)=\left. \frac{3}{4\pi}
(\lambda-sin\lambda cos\lambda)(\frac{1}{3}f_{n}^{3}-f_{n})
\right|^{f_{n}(\infty)=-1}_{f_{n}(0)=1}.                \label{l5}
\ee
Noting that the gauge potential (\ref{l9}) vanishes  at  $\lambda=0$, 
one obtains the following value of the  CS  number  along  the  path 
(\ref{l9}) \cite{sphal}:
\be
N_{CS}=\left.\int\sqrt{-g}K^{0}d^{3}x\right|_{t}=
\frac{1}{\pi}(\lambda-sin\lambda cos\lambda),            \label{l12}
\ee
which changes from zero to one as $\lambda$ runs from $0$  to  $\pi$. 

Thus the fields (\ref{2}), (\ref{l8}), (\ref{l9}) interpolate  between 
two EYM vacua with different winding numbers of the gauge field, and 
coincide with  the  (gauge  transformed)  field  of  the  $n$-th  BK 
solution when $\lambda=\pi/2$. This allows one to treat  odd-$n$  BK 
solutions as sphalerons, ``lying''  on  the  top  of  the  potential 
barrier  separating  distinct  topological  vacuum  sectors  of  the 
theory.  The  profile  of  this  barrier  may   be   obtained   from 
Eq.(\ref{l8}) due to the fact that the metric function $m(r)$  obeys
the  initial value  constraint,  which  allows one  to  define  the
ADM   energy \cite{sphal} \cite{sphal}, \cite{small}
$$ 
E[\lambda,f(r)] = \lim_{r\rightarrow\infty}m(r)=
$$
\be
=sin^{2}\lambda\iii (f'^{2} + 
sin^{2}\lambda\frac{(f^{2}-1)^{2}}{2r^{2}})
exp(-2sin^{2}\lambda\int_{r}^{\infty}f'^{2}\frac{dr}{r})dr,
                                                         \label{l13} 
\ee
such that the sphaleron energy is given by $E[\pi/2,f_{n}(r)]$.

\section{Perturbations of the equilibrium solutions}

The functional $E$ given by (\ref{l13}) is quite useful for a
qualitative understanding of the behaviour of the potential
barrier surface in the vicinity of sphalerons \cite{small}.
The sphaleron configurations, $\lambda=\pi/2, f(r)=f_{n}(r)$,
correspond to critical points of $E[\lambda,f(r)]$. 
One can easily see that $E[\lambda,f(r)]$ reaches its
maximal value at $\lambda=\pi/2$, which shows that each
sphaleron possesses an unstable mode \cite{sphal};
this kind of instability can be naturally called sphaleron
instability. 
In addition, the value of $E[\pi/2,f_{n}(n)]$ decreases
under certain deformations of another type: $f_{n} \rightarrow
f_{n}+\delta f$ \cite{small} which will be referred to as
gravitational instabilities.

To put the this into a more rigorous form, we return to the general
system of the EYM field equations (\ref{6})-(\ref{10}) and
consider small spherically-symmetric perturbations of a
given (regular or black hole) equilibrium solution:
\be
m\rightarrow m+\delta m,\ \
\sigma\rightarrow\sigma +\delta\sigma,\ \
f\rightarrow f+\delta f,\ \
W_{0}=W_{1}=0\rightarrow \delta W_{0},\ \delta W_{1}.    \label{11}
\ee
In view of (\ref{5:1}), these perturbations
can be classified  in  accordance  with 
their parity. Observing that the  background  equilibrium  solutions 
are  invariant  under   parity,   one concludes   that 
even--parity   perturbations,  $(\delta   m,\delta\sigma,     \delta 
p_{1})$,   and  odd--parity    perturbations,    $(\delta     W_{0}, 
\delta   W_{1},  \delta  p_{2})$,    must be independent,  which  is 
confirmed      by the  straightforward       linearization       of 
Eqs.(\ref{6})-(\ref{10}). Notice   that   the  infinitesimal   gauge 
transformation     (\ref{5})   does   not  alter   the   even-parity 
perturbations, while the odd-parity ones change as 
\be
\delta W_{0}\rightarrow\delta W_{0}+\partial_{t}\Omega, \ \ \
\delta W_{1}\rightarrow\delta W_{1}+\partial_{r}\Omega, \ \ \
\delta p_{2}\rightarrow\delta p_{2}+\Omega f.            \label{12}
\ee

Perturbations  in  the  even-parity  sectors   have   been   studied 
in   \cite{SZ}.   In   this   case,   linearization   of 
Eqs.(\ref{7}), (\ref{9}), (\ref{10}) reveals that  perturbations  of 
the  metric  functions,  $\delta  m$  and  $\delta\sigma$,  can   be 
expressed entirely in terms of $\delta p_{1}$, the latter  satisfies 
the following Schr\"odinger-type equation: 
\be
-\phi''+\left(\sigma^{2}N\frac{3f^{2}-1}{r^{2}}+
2\left(\frac{\sigma'}{\sigma}\right)'\right)\phi
=\omega^{2}\phi,                                        \label{12:1}
\ee
where $\delta  p_{1}=\exp(-i\omega  t)\phi(\rho)$ with $\rho$
being the tortoise coordinate introduced above
(throughout  this
paper  we  use   the   following   notations:   $y'=dy/d\rho$,   and 
$y'_{r}=dy/dr$). For the $n$-th background  regular  or  black  hole 
solution, numerical analysis has shown the existence  of  $n$  bound 
states in (\ref{12:1}).

The even parity perturbations correspond to the
gravitational perturbations defined above. 
It is instructive to see how
the perturbation equation (\ref{12:1}) can be derived directly
from the potential barrier functional $E$ \cite{small}.
Let us introduce  {\it a barrier height  functional}
\be
\varepsilon[f(r)]=E[\lambda=\pi/2,f(r)],              \label{s49}
\ee 
and consider its expansion near a critical point
\be 
\varepsilon [f_{n}(r)+\phi(r)]=
\varepsilon[f_{n}(r)]+\delta^{2}\varepsilon +\ldots , 
                                                       \label{s52}
\ee
where the first order term vanishes and dots denote higher order 
terms. The second order term reads 
\be
\delta^{2}\varepsilon = \iii\phi
(-\frac{d^{2}}{d\rho^{2}}  +V)\phi d\rho,             \label{s53}
\ee
The effective potential $V$ in this expression is explicitly
derived in the Appendix, and turns out to be the same as
that in the Schr\"odinger equation (\ref{12:1}). This means
that any solution of the perturbation problem (\ref{12:1})
with $\omega^{2}<0$ gives rise to the negative second
variation of the barrier height functional.

Let us pass now to the analysis of the odd-parity 
modes. The odd-parity perturbation sector contains the
sphaleron instabilities of the equilibrium solutions.
For the odd-parity perturbations,
the  metric  remains  unperturbed, such that
the  perturbation 
equations can  easily  be  obtained  via  expanding  the  Yang-Mills  
equations   (\ref{6}),   (\ref{7})   with   respect    to    $\delta   
W_{0}$,   $\delta   W_{1}$,    and    $\delta p_{2}$ alone.  We  use  
the  gauge  freedom  (\ref{12})   to  impose the condition $\delta 
W_{0}=0$,  and   specify   the   time    dependence  as      $\delta     
W_{1}=\exp(-i\omega      t)\alpha(r)$,      $\delta  p_{2}=\exp(-\ii 
\omega t)\xi(r)$. The resulting  equations read 
\be
\left( -\frac{d}{dr}N\sigma\frac{d}{dr}+
\frac{\sigma}{r^{2}}(f^{2}-1)\right)\xi+
(f_{r}'+\frac{d}{dr}f)N\sigma\alpha
=\frac{\omega^{2}}{\sigma N}\xi,                      \label{13}
\ee
\be
2N\frac{\sigma^{2}}{r^{2}}\left( f^{2}\alpha+f_{r}'\xi -
f\xi_{r}'\right)=\omega^{2}\alpha,                       \label{14}
\ee
where $f$, $\sigma$ and $N$  refer  to  the  background  equilibrium 
solution. There exists also an additional equation due to the  Gauss 
constraint (Eq.(\ref{6}) with $\nu=0$): 
\be
\omega(\left(\frac{r^{2}}{\sigma}\alpha\right)_{r}'-
\frac{2f}{N\sigma}\xi) =0.                            \label{15}
\ee
To simplify  these   equations  we  introduce   the   new   variable  
$\chi= (r^{2}/2\sigma)\alpha$ and pass to the  tortoise   coordinate  
$\rho$. Multiplying (\ref{13}) 
by $f$ and using the background  Yang-Mills equation 
\be
f''=\frac{\sigma^{2}N}{r^{2}}f(f^{2}-1),               \label{16}
\ee
one can represent Eqs.(\ref{13}-\ref{15}) in the following form:
\be
\left( f^{2}\gamma^{2}\chi+f'\xi-f\xi'\right)'= 
\omega^{2}f\xi,                                         \label{17}
\ee
\be
 f^{2}\gamma^{2}\chi+f'\xi-f\xi'=\omega^{2}\chi,         \label{18}
\ee
\be
\omega (\chi'-f\xi)=0,                                  \label{19}
\ee
where $\gamma^{2}=2N\sigma^{2}/r^{2}$. One can immediately see that, 
as  long  as  $\omega\neq  0$,  only  two  of  these  equations  are 
independent,   say,   (\ref{18})   and   (\ref{19})   (one  can 
equally say that Eqs.(\ref{17}), (\ref{18}) are independent, and  the   
Gauss constraint  in  the  case   is   the differential
consequence   of   this  
equation    of   motion).     When     $\omega=0$,     Eq.(\ref{19})   
disappears,    and  there  remains  only  one  essential   equations 
(\ref{18}) which can be solved by any static  pure   gauge  obtained 
from (\ref{12}): 
\be
\chi=\frac{\Omega'}{\gamma^{2}},\ \ \  
\xi=f\Omega.                                              \label{20}
\ee
Let $\omega$ be non-vanishing. Then one can omit the factor $\omega$ 
in (\ref{19}) and  specify  the  following  independent   equations 
which are clearly equivalent to (\ref{17})-(\ref{19}): 
\be
f\xi'-f'\xi=(f^{2}\gamma^{2}-\omega^{2})\chi,             \label{21}
\ee
\be
\chi'=f\xi.                                               \label{22}
\ee
When $\omega=0$,  Eq.(\ref{22})  does  not  appear  in  the  initial 
system, however we will retain this equation at zero energy too.  In 
this case, this equation (it  is  sometimes  called  ``strong  Gauss 
constraint'') plays the role of the gauge  fixing  condition.  Imposing 
this  condition  violates  the  remaining  gauge    invariance    of  
equations (\ref{17})-(\ref{19}) at zero  energy,  allowing  only for 
those pure gauge  solutions  (\ref{20})  whose  parameter   $\Omega$  
satisfies 
\be
\left(\frac{\Omega'}{\gamma^{2}}\right)'=f^{2}\Omega.  \label{23}
\ee

\section{Transformation to Schr\"odinger problem}

Our aim now is to   reduce  Eqs.(\ref{21}),    (\ref{22})    to    a 
one-channel Schr\"odinger problem. The most natural way to  do  this 
is to exclude  one  of  the  functions, either   $\xi$   or  $\chi$, 
from the  equations.  In  this  case  however,  one  obtains  either 
singular   or   energy-dependent   effective  potential    in    the    
resulting Schr\"odinger equation. For instance, eliminating
$\xi$ one finds
\be
-\phi''+
\left(\gamma^{2}(f^{2}+1)+2\left(\frac{f'}{f}\right)^{2}\right)
\phi
=\omega^{2}\phi,                                        \label{g16}
\ee
where $\phi=\chi/f$, the potential in this equation is ill-defined
due to nodes of $f$.
Another possibility is to  try  to
look for  a certain linear combination of  $\chi$  and  $\xi$  which 
might satisfy  a  Schr\"odinger-type    equation    with    regular,
energy-independent potential. Surprisingly, this way turns out to  be
successful. The key observation here is that the combination
\be
\psi=fZ\chi+\xi,                                      \label{24}
\ee
where $Z$ is a solution of the following auxiliary equation:
\be
Z'=f^{2}Z^{2}-\gamma^{2},                             \label{25}
\ee
satisfies by virtue of (\ref{21}), (\ref{22}) and (\ref{25}) 
the Schr\"odinger-type equation 
\be
-\psi''+U\psi=\omega^{2}\psi,                         \label{26}
\ee
with the energy-independent effective potential
\be
U=\sigma^{2}N\frac{3f^{2}-1}{r^{2}}+
2\left( f^{2}Z\right)'.                               \label{27}
\ee
The inverse transformation from  $\psi$  to  the 
variables $\chi$ and $\xi$ reads
\be
\chi=\frac{1}{\omega^{2}}\left( f^{3}Z\psi+f'\psi-f\psi'\right),
\ \ \ \ \ \ \ \ \ \ \xi=\psi-fZ\chi.                     \label{28}
\ee
In the new potential derived the nodes of $f$ are no more
dangerous, such that the potential is regular provided that
$Z$ is regular. Furthermore, the investigation of the behaviour of $Z$
is simplified considerably due the existence of 
another remarkable relation allowing us
to    represent    solutions     of
Eq.(\ref{25})  in  terms  of   solutions    of   second-order   {\sl 
linear} differential equation (\ref{23}): 
\be
Z=-\Omega\Lambda-\frac{\Lambda^{2}}
{C+{\displaystyle\int_{0}^{\rho}}f^{2}\Lambda^{2}d\rho}, 
\ \ \ \ {\rm where}\ \ \ \Lambda=\frac{\gamma^{2}}{\Omega'}, 
                                                        \label{28:1}
\ee
with $C$  being  an  integration  constant.  For   each   particular 
$\Omega(\rho)$, this correspondence provides us  with  a  family   of  
solutions to Eq.(\ref{25}) whose  members   are   distinguished   by 
values of $C$ in (\ref{28:1}). However, not all of  these  solutions 
are well-behaved.  Our  aim  is  to  specify  $\Omega$  and  $C$  in 
(\ref{28:1}) such that  the   resulting   $Z(\rho)$  gives  rise  to 
well-defined  potential (\ref{27}) and  transformations  (\ref{24}), 
(\ref{28}).

Before passing to the further analysis, one may wonder why 
it was possible to guess the transformation presented. 
One might suspect the existence of some symmetry in the problem,
which could naturally explain the appearance of all the 
relations listed above. It turns out that there 
exists indeed a regular way to proceed in the case, since 
the transformation applied formally resembles
the transformation between the dual partners in
supersymmetric quantum mechanics \cite{super}.
First, notice that Eq.(\ref{g16}) has obvious $\omega=0$
solutions which are just pure gauge modes (see (\ref{20})),
\be
\phi_{0}=\Omega'/f\gamma^{2},                          \label{g17}
\ee
with $\Omega$ satisfying (\ref{23}). 
Any such solution allows us to factorize the differential
operator in (\ref{g16}):
\be
-\frac{d^{2}}{d\rho^{2}}+
\left(\gamma^{2}(f^{2}+1)+2\left(\frac{f'}{f}\right)^{2}\right)=
-\frac{d^{2}}{d\rho^{2}}+\frac{\phi_{0}''}{\phi_{0}}
=Q^{+}Q^{-},                                            \label{g19}
\ee
with
\be
Q^{\pm}=\mp\frac{d}{d\rho}-\frac{\phi_{0}'}{\phi_{0}}. \label{g20}
\ee
Using (\ref{23}) we find
\be
Q^{\pm}=\mp\frac{d}{d\rho}-\frac{f'}{f}-f^{2}Z,         \label{g21}
\ee
where
\be
Z=-\Omega\Lambda,\ \ \ \Lambda=\gamma^{2}/\Omega'.   \label{g22}
\ee
Notice that (\ref{23}) is equivalent to
\be
\Lambda'/\Lambda=f^{2}Z,                             \label{g23}
\ee
and  thus $Z$ satisfies the nonlinear differential equation
(\ref{25}).

With  a standard reduction, the most general solution of the
second order linear equation (\ref{23}) is
\be
\Omega=c_{2}\tilde{\Lambda}+\tilde{\Omega}
\left(c_{1}+c_{2}\int_{0}^{\rho}f^{2}
\tilde{\Lambda}^{2}d\rho\right),                    \label{g25}
\ee
where $\tilde{\Omega}$ is a special solution, $\tilde{\Lambda}=
\gamma^{2}/\tilde{\Omega}'$, and $c_{1}$, $c_{2}$ are real
constants. This gives immediately for $Z$ in (\ref{g22})
\be
Z=-\tilde{\Omega}\tilde{\Lambda}-
\frac{\tilde{\Lambda}^{2}}
{c_{1}/c_{2}+{\displaystyle\int_{0}^{\rho}}f^{2}
\tilde{\Lambda}^{2}d\rho},                           \label{g26}
\ee
which is obviously identical to (\ref{28:1}). 
Since the differential equation (\ref{g16}) is, for any $\Omega$
in (\ref{g25}), identical to
\be
Q^{+}Q^{-}\phi=\omega^{2}\phi,                       \label{g27}
\ee
one can pass from $\phi$ to
\be
\psi=Q^{-}\phi.                                      \label{g28}
\ee
For $\omega\neq 0$ this has the unique inverse
\be
\phi=\frac{1}{\omega^{2}}Q^{+}\psi                   \label{g29}
\ee
and by applying $Q^{-}$ we obtain the ``dual'' eigenvalue
equation
\be
Q^{-}Q^{+}\psi=\omega^{2}\psi.                       \label{g30}
\ee
The differential operator on the left is 
\be
Q^{-}Q^{+}=-\frac{d^{2}}{d\rho^{2}}+U,              \label{g31}
\ee
with $U$ being given by (\ref{27}), whereas transformations
(\ref{g28}), (\ref{g29}) are identical to (\ref{24}), (\ref{28}). 

\section{Structure of the potential}

Now we turn to the investigation of the behaviour of $Z$.
First,  consider  the  case  of  the regular background equilibrium 
solutions.  It  is  convenient  to  return  for  a  moment  to   the  
variable $r$, then Eq.(\ref{23}) takes the form 
\be
r^{2}\Omega_{rr}''+2r(1-f_{r}'^{2})\Omega_{r}'=
\frac{2f^{2}}{N}\Omega.                           \label{28:2}
\ee
This  differential  equation  has   regular   singular   points   at 
$r=0,\infty$. As  is  known  \cite{BFM},  the  background  solutions 
(\ref{a1}), (\ref{a2}) are analytic in  some  neighborhoods  of  the 
origin and infinity thus ensuring analyticity of the coefficients in 
(\ref{28:2}). This  guarantees  the  existence  of  the  fundamental 
system  of  solutions  in  the  vicinity  of  each  singular  point 
\cite{WW}.  Using  (\ref{a1}),  (\ref{a2}),  (\ref{28:2})  one   can 
represent these solutions in the following form: 
\be \displaystyle
\Omega (r)=c_{1}~r\left\{ 1+R_{1}(r)\right\} +
c_{2}~\frac{1}{r^{2}}\left\{ 1+R_{2}(r)\right\} \ \ \ \
{\rm as}\ \ r\rightarrow\ 0,                           \label{28:3}
\ee
and introducing $s=1/r$,
\be
\Omega (s)=d_{1}~s^{2}\left\{ 1+S_{1}(s)\right\} +
d_{2}~\frac{1}{s}
\left\{ 1+S_{2}(s)+{\cal A}~s^{3}ln(s)(1+S_{1}(s))\right\} \ \ \ \
{\rm as}\ \ s\rightarrow\ 0,               \label{28:4}
\ee
where   $c_{1,2}$   and   $d_{1,2}$   are    arbitrary    constants,  
$R_{1,2}(r)$ and $S_{1,2}(s)$ are convergent power series  vanishing 
at  $r=0$ and $s=0$, respectively,
the quantity  ${\cal  A}={\cal   A}(a,M)$    is  
specified  by  the  asymptotic behaviour of the background solution. 

In the domain $0<r<\infty$ Eq.(\ref{28:2}) has  no  singular  points 
and the coefficients of the   equation   are   at  least  continuous 
functions in this region.  This  guarantees   the  existence  of  an 
extension of the asymptotic solutions (\ref{28:3}), (\ref{28:4})  to 
the whole half-line, and this extension  is at least $C^{2}$. 

Now we choose  a special  solution  $\tilde{\Omega}$  which  is
regular  at  infinity. For this we  specify  in  (\ref{28:4}) 
$d_{1}=1$,  $d_{2}=0$.  Then,  near  the   origin,  $\tilde{\Omega}$ 
is given by (\ref{28:3}).  One   can   see   that   the  coefficient 
$c_{2}$ in this formula  does   not   vanish.   Indeed,  multiplying 
Eq.(\ref{23})  by $\Omega$ and integrating  from  $\rho$ to infinity 
one obtains 
\be
\left.
0=\int_{\rho}^{\infty}(-\Omega\left(\frac{\Omega'}{\gamma^{2}}
\right)'+f^{2}\Omega^{2})d\rho = 
- \frac{\Omega'}{\gamma^{2}}\Omega\right|_{\rho}^{\infty}+
\int_{\rho}^{\infty}(\frac{\Omega'^{2}}{\gamma^{2}} +
f^{2}\Omega^{2})d\rho.                            \label{30}
\ee
Applying this relation to the particular solution  $\tilde{\Omega}$, 
$\tilde{\Omega}(\infty)=\tilde{\Omega}'(\infty)=0$,
one arrives at 
\be
(\tilde{\Omega}^{2})'=-2\gamma^{2}
\int_{\rho}^{\infty}(\frac{\tilde{\Omega}'^{2}}{\gamma^{2}} +
f^{2}\tilde{\Omega}^{2})d\rho < 0,                    \label{31}
\ee
so that, since $\tilde{\Omega}$ vanishes at  infinity,  it  can  not 
vanish  also at the origin containing therefore the growing  branch, 
$\tilde{\Omega}\sim 1/r^{2}$  as  $r\rightarrow  0$.  From  here  we 
conclude    that on the domain $0<r<\infty$ there exists  a   smooth 
(at  least  $C^{2}$)   solution   $\tilde{\Omega}(r)$   with     the   
following   behaviour    near    $r=0,\infty$: 
\be
\tilde{\Omega}= \frac{const}{r^{2}}+O(1)\ \ 
{\rm as}\ r\rightarrow 0,\ \ \ \ \ \ \ \ \ \ 
\tilde{\Omega}= \frac{1}{r^{2}}+O(\frac{1}{r^{3}})
\ \ {\rm as}\ r\rightarrow\infty.          \label{32}
\ee

Next, we substitute this solution    in    Eq.(\ref{28:1}),    where 
we require the constant $C$ to be positive -- to avoid vanishing  of 
the  denominator.   Then,    for    each  $C>0$,    the    resulting  
solution  $Z(r)$ is   smooth   and differentiable  in  the  interval 
$(0,\infty)$   with    the   following behaviour near the origin and 
infinity: 
$$
Z= \sigma_{0}\left( \frac{1}{r}+\ldots+p(C)r^{2}+O(r^{3})\right)
\ \ \ \ \ {\rm as}\ \ \  r\rightarrow 0,\ 
$$
\be
Z= -\frac{2}{r^{2}}+\ldots+\frac{q(C)}{r^{4}}+O(\frac{ln(r)}{r^{5}})
\ \ \ \ \ {\rm as}\ \ \ r\rightarrow\infty.             \label{33}
\ee
Here  dots  denote  terms  specified  entirely  by  the   background 
solutions (\ref{a1}), (\ref{a2}), whereas the quantities $p(C)$  and 
$q(C)$ depend on the constant $C$ as well. Using (\ref{a4})  we  can 
return back to the tortoise coordinate $\rho$ thus obtaining 
\be
Z= \frac{1}{\rho}+O(\rho)\ \ {\rm as}\   \rho\rightarrow 0,
\ \ \ {\rm and}\ \ 
Z= -\frac{2}{\rho^{2}}+O(\frac{ln(\rho)}{\rho^{3}})
\ \ {\rm as}\  \rho\rightarrow\infty.                \label{33:1}
\ee
We therefore arrive at the one-parameter family of solutions  $Z(\rho)$, 
whose members are distinguished by values of  the  constant  $C>0$  in 
(\ref{28:1}),   each   solution    satisfies   boundary    condition 
(\ref{33:1}).   Some   of   the   typical solutions are presented in 
Fig.1. All these  solutions have only one  zero  in  the
interval  $(0,\infty)$  --   this   is    because,    in   view   of  
Eq.(\ref{25}),  the   derivative  $Z'$  is   always  negative   when  
$Z$  vanishes. One   should  stress  that  such a behaviour of   $Z$  
is common for all  background  regular equilibrium solutions. 


\begin{figure}
\epsfxsize=8cm
\centerline{\epsffile{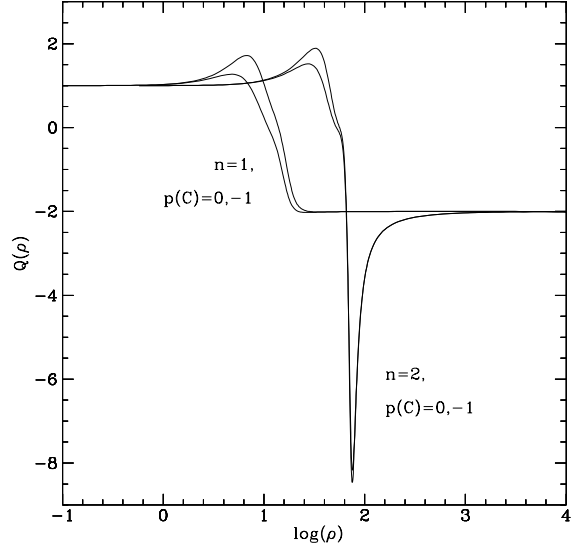}}
\caption{Behavior of  the  function  $Z(\rho)$  for  $n=1$  and  $n=2$ 
background BK solutions. Instead of $Z$, it is convenient to display 
$Q$  such  that  $Z=\sigma Q/r$.  For  each  $n$,  a  pair  of 
solutions is shown corresponding to the two different choices of the 
integration constant $p(C)$ in Eq.(70)  (the  higher  maximums 
correspond to $p(C)=0$). }
\label{Fig.1}
\end{figure}

Now, having  obtained  the  function  $Z$,  we   can  determine  the 
effective  potential  $U(\rho)$  in   the   Schr\"odinger   equation 
(\ref{26}).  Notice  that   the   first term      in      (\ref{27})     
contains              an              unbounded               piece: 
$\sigma^{2}N(3f^{2}-1)/r^{2}\rightarrow   2/\rho^{2}$   when   $\rho 
\rightarrow 0$. However, in account of condition  (\ref{33:1}), this   
divergency is   exactly   canceled   by   the   second    term    in 
(\ref{27}), such that the resulting  potential  turns  out   to   be  
continuous  in  the  whole semi-axis, $\rho\in [0,\infty)$, with the 
following behaviour near the origin and infinity: 
\be
U(\rho)=6b(6b-1)\sigma_{0}^{2}+O(\rho)\ \ {\rm as}
\ \ \rho\rightarrow 0,\ \
U(\rho)=\frac{6}{\rho^{2}}+O(\frac{ln(\rho)}{\rho^{3}})\ \ 
{\rm as}\ \ \rho\rightarrow\infty,                       \label{34}
\ee
where $b$ comes from  Eq.(\ref{a1}).  We therefore  find   a  whole  
family  of  regular  potentials.   Each  member  of   this   family, 
$U(\rho)$, is specified by the  corresponding  solutions   $Z(\rho)$  
and   satisfies boundary conditions (\ref{34}). A typical example of 
such a potential is depicted in Fig.2.


\begin{figure}
\epsfxsize=8cm
\centerline{\epsffile{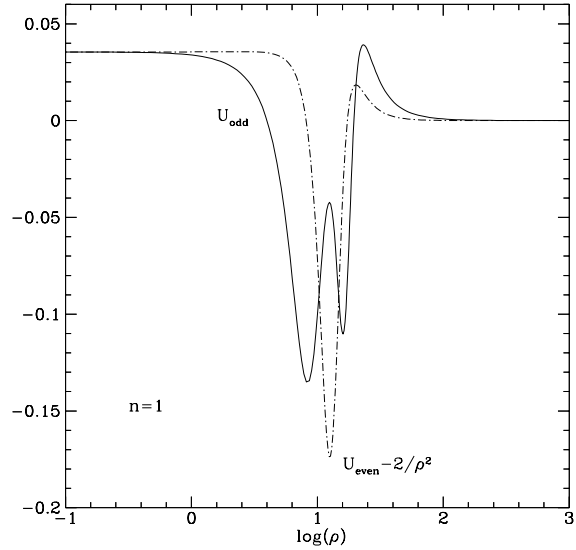}}
\caption{
Odd-parity and even-parity potentials for the $n=1$ background 
BK solution. The odd-parity potential is given by Eq.(48), the 
corresponding $Z$ is shown in Fig.1 for  $p(C)=0$.  The  even-parity 
potential is defined by Eq.(29). 
}
\label{Fig.2}
\end{figure}

Thus the potential  $U$  in  the Schr\"odinger equation
(\ref{26}) indeed can be chosen 
to  be  regular 
everywhere in  the  half-line  $[0,\infty)$.  However,  $U$  is  not 
uniquely defined. Our aim now is to show that all  potentials  from 
the family specified above are equivalent as far as the bound  state 
problem is concerned. To this end we study  the  properties  of  the 
transformations (\ref{24}), (\ref{28}). 

Consider a solution of the initial system of  equations  (\ref{21}), 
(\ref{22}) corresponding to a bound state with $\omega^{2}<0$. Then, 
near the origin and infinity, 
\be
\xi\sim\rho\ ,\ \ \ \chi\sim\rho^{2}\ \ \ {\rm as}\ 
\rho\rightarrow 0,\ \ \ \ \ \ \ \ \ \ \ \
\xi\sim\pm\frac{1}{\omega}~\chi\sim exp(-|\omega|\rho)
\ \ \ {\rm as}\ \ \rho\rightarrow\infty.               \label{34:1}
\ee
Applying transformation (\ref{24}) with  some  particular  $Z(\rho)$ 
and taking (\ref{33:1}) into account one obtains the solution  of 
Eq.(\ref{26}) with the following behaviour: 
\be
\psi\sim\rho\ \ {\rm as}\ \ \rho\rightarrow 0,\ \ \ \ \ \ \  
\psi\sim exp(-|\omega|\rho)\ \ {\rm as}
\ \ \rho\rightarrow \infty ,                 \label{34:2}                   
\ee
which ensures normalizability of $\psi$  
\be
\langle\psi|\psi\rangle =\int_{0}^{\infty}
\psi^{2}~d\rho < \infty ;                       \label{34:3}
\ee
so that $\psi$ corresponds  to  the  bound  state  solution  of  the 
Schr\"odinger equation (\ref{26}) with the same  $\omega^{2}<0$.  On 
the other hand, starting from (\ref{34:2}) and applying the  inverse 
transformation (\ref{28}), one arrives at (\ref{34:1}). It is  worth 
noting that these transformations uniquely fix the  value  $\psi(0)=0$ 
ensuring essential self-adjointness of the Schr\"odinger    operator   
$H=-d^{2}/d\rho^{2}+U(\rho)$ (the Schr\"odinger equation  by  itself 
does not specify $\psi(0)$ because the potential is globally  
regular in the case).

Thus,  for  each   particular   $U$,   transformations   (\ref{24}), 
(\ref{28}) establish one-to-one correspondence between bound  states 
with non-zero energy for the initial system of equations (\ref{21}), 
(\ref{22}) and those  for  the  Schr\"odinger  equation  (\ref{26}). 
Obviously, this means that the different choices  of  the  potential 
$U$ are  equivalent.  Precisely,  the  nature  of  this 
equivalency is explained by the following  considerations.  
Let  $Z_{1}$ 
and  $Z_{2}$  be   two   different   solutions   of    Eq.(\ref{25}) 
(relating  of    course    to    the    same    background     field 
configuration),  and   $U_{1}$,  $U_{2}$   are   the   corresponding 
potentials. Then, given a solution of (\ref{26})  for  the 
potential $U_{1}$   and   some   $\omega^{2}\neq  0$,    $\psi_{1}$,   
one  can construct a solution for  the   potential    $U_{2}$ and  
{\sl the same} $\omega^{2}$ due to the following rule: 
\be
\psi_{2}=\psi_{1}+\frac{1}{\omega^{2}}f(Z_{2}-Z_{1})
\left( f^{3}Z_{1}\psi_{1}+f'\psi_{1}-f\psi_{1}'\right).   \label{35}
\ee
Suppose that $\psi_{1}$  is  a  normalizable solution  corresponding  
to a bound state of $U_{1}$, then it  satisfies  (\ref{34:2}).  From  
(\ref{33}), (\ref{a4}) one concludes that 
\be
Z_{1}-Z_{2}\sim\rho^{2}\ \ {\rm as}\ \ \rho\rightarrow 0, 
\ \ \ \ \ \
Z_{1}-Z_{2}\sim\frac{1}{\rho^{4}}
\ \ {\rm as}\ \ \rho\rightarrow\infty,                  \label{35:1}           
\ee
which ensures that $\psi_{2}$ also satisfies (\ref{34:2}), that  is, 
it  relates to a bound state of $U_{2}$  with  the  same  eigenvalue 
$\omega^{2}<0$. 
Thus the bound  state  problem  for  Eqs.(\ref{21}),  (\ref{22})  is 
equivalent to that for the Schr\"odinger  equation  (\ref{26})  with 
any of the effective potentials from the family  specified  above.

\section{Number of bound states}

Our 
purpose now is to show that Eq.(\ref{26}) admits a zero energy bound 
state. The simplest way to see this is to start from Eqs.(\ref{21}), 
(\ref{22}) which are known to possess the pure gauge solutions  with 
zero energy. These solutions are given  by  Eq.(\ref{20})  with  the 
gauge   parameter   $\Omega$   satisfying    Eq.(\ref{23}).    Then, 
substituting (\ref{20}) in (\ref{24})  one  obtains  a  zero  energy 
solution    for    Eq.(\ref{26}),        $\psi_{0}=f\phi$,    where 
$\phi=\Omega+Z\Omega'/\gamma^{2}$. Next, observing that the function 
$\phi$ obeys the following  equation,  $\phi'=f^{2}Z\phi$,  one  can 
represent $\psi_{0}$ as 
\be
\psi_{0}=f\exp\left(\int_{\rho_{0}}^{\rho}f^{2}Zd\rho\right),
                                                     \label{36}
\ee
with some positive $\rho_{0}$; the validity  of  this  solution  can 
also be checked directly. The exponential factor in this  expression 
is manifestly positive when $0<\rho<\infty$ and, on account  of  the 
boundary conditions for $Z$, vanishes at $\rho=0,\infty$, such that 
\be
\psi_{0}\sim\rho\ \ {\rm as}\ \ \rho\rightarrow 0, 
\ \ \ \ \ \ \ \ \
\psi_{0}\sim\frac{1}{\rho^{2}} 
\ \ {\rm as}\ \ \rho\rightarrow\infty,  \label{37} 
\ee
which   ensures  normalizability   of   $\psi_{0}$.   Thus 
$\psi_{0}$,  being  a  normalizable  solution   of   (\ref{26})   and 
satisfying   the   right   boundary   condition   at   the   origin, 
$\psi_{0}(0)=0$, relates in fact to zero energy bound state. One 
should stress that this particular bound state occurs only  for  the 
Schr\"odinger equation (\ref{26}) and not for the initial system  of 
equations (\ref{21}), (\ref{22}).  In fact, the equivalence  between 
the initial system and the Schr\"odinger equation  holds   only  for 
non-zero energy.  When  $\omega^{2}=0$,  transformation   (\ref{28}) 
becomes ill-defined, and the equivalence is lost. Then, taking  into 
account Eq.(\ref{32})  one can see that  the  pure  gauge  solutions 
(\ref{20}) by themselves can be neither normalizable nor  even  just 
regular both at the origin  and  infinity.  However,  their  certain 
linear combination, $\psi_{0}$, can be normalizable. The idea is  to 
use $\psi_{0}$ to establish the existence of {\sl  negative  energy} 
bound states for the Schr\"odinger equation, and those are certainly 
equivalent to bound states appearing in the initial system. 

To this end we observe that, owing to the oscillating factor $f$  in 
(\ref{36}), $\psi_{0}$ has exactly $n$ nodes, which  clearly  proves 
that the Schr\"odinger equation admits  $n$  bound  state  solutions 
with negative energies.  We  therefore  conclude  that  
for the $n$-th regular BK solution there exist
precisely  $n$  negative modes in the odd-parity spherical 
perturbation sector.

\section{Black hole case}

For black holes, the proof is essentially the same as  that  in  the 
regular case. We start again form Eq.(\ref{23}) and want to  specify 
a particular solution  for  this  equation,  $\tilde{\Omega}(\rho)$, 
such that the corresponding $Z(\rho)$ is well behaved. The  variable 
$\rho$  now  runs   over   the   line   $-\infty   <\rho   <\infty$, 
such that
$\rho=-\infty$ (horizon) and  $\rho=+\infty$  are  regular  singular 
points of the differential equation. Instead of $\rho$, 
it is convenient to use the independent variable $x$ defined by
(\ref{a3:1}), then  Eq.(\ref{28:2}) takes the form 
\be
x(1+x)^{2}~\Omega_{xx}''+
2x(1+x)\left(1-\left(f_{x}'\right)^{2}/r_{h}^{2}\right)\Omega_{x}'=
2f^{2}\frac{x}{N}~\Omega.                           \label{38}
\ee
In this equation, the coefficients are  analytic  functions  in  the 
vicinity of $x=0$, because the  background  solutions  (\ref{a3})  are 
analytic in some  neighborhood  of  the  horizon  \cite{BFM}.  This 
ensures the existence of the fundamental system of solutions in  the 
vicinity   of   $x=0$.   We   choose   the    particular    solution 
$\tilde{\Omega}$ whose regular integral vanishes at the horizon: 
\be
\tilde{\Omega}(x)=x(1+X(x)),\ \ \ {\rm as}
\ \ \ x\rightarrow 0,                              \label{39}
\ee
where $X(x)$ is given by a power series convergent in  the  vicinity 
of $x=0$, $X(0)=0$. Since the  coefficients  in  (\ref{38})  are  at 
least continuous in the domain $x>0$, the solution (\ref{39}) admits 
at a smooth extension  into  this  region.   When  $x$  tends  to 
infinity, the solution is given by (\ref{28:4}) where, repeating the 
arguments used in Eqs.(\ref{30}), (\ref{31}), the coefficient before 
the growing part can be shown to be  non-vanishing.  
From here one concludes 
that on the domain $0<x<\infty$ there exists a smooth  solution 
of (\ref{38}) with the following behaviour near $x=0,\infty$: 
\be
\tilde{\Omega}= x+O(x^{2})\ \ 
{\rm as}\ x\rightarrow 0,\ \ \ \ {\rm and}\ \ 
\tilde{\Omega}= const~x+O(1)
\ \ {\rm as}\ x\rightarrow\infty.                  \label{40}
\ee
Next, using (\ref{a4}), (\ref{a5}) we express this solution in terms 
of $\rho$ and substitute it into Eq.(\ref{28:1}), where, in  order  to 
ensure the non-vanishing of the denominator, we choose the  constant 
$C$ such that 
\be
C<-\int_{0}^{\infty}f^{2}\Lambda^{2}d\rho;      \label{41}
\ee
the integral here exists since Eq.(\ref{40}) ensures that
$\Lambda\sim 1/\rho^{2}$ when
$\rho\rightarrow \infty$. 
For  each  certain  $C$  we  obtain  from  (\ref{28:1})  a  solution 
$Z(\rho)$ which is continuous and differentiable 
in $(-\infty,+\infty)$ with the 
following behaviour near the horizon and infinity: 
\be
Z= -\frac{1}{f^{2}_{h}\rho}+O(\frac{1}{\rho^{2}})\ \ {\rm as}\ \
\rho\rightarrow -\infty,\ 
\ \ \ {\rm and}\ \ 
Z\rightarrow -\frac{2}{\rho^{2}}+O(\frac{ln(\rho)}{\rho^{3}})
\ \ {\rm as}\ \
\rho\rightarrow\infty.                                  \label{42}
\ee
The  corresponding  effective  potential  $U(\rho)$  is  smooth  and 
bounded. The equivalency of the different particular $U$ can be
checked with the use of (\ref{35}), (\ref{42}).
Then,  repeating  the  analysis  performed  above  for  the 
regular case, we arrive at the zero energy solution $\psi_{0}$ given 
by (\ref{36}). In view of (\ref{42}), this  function  satisfies  the 
following boundary conditions: 
\be
\psi_{0}\sim\frac{1}{\rho}\ \ {\rm as}\ \ \rho\rightarrow -\infty, 
\ \ \ \ \ \  \ \
\psi_{0}\sim\frac{1}{\rho^{2}}\ \ {\rm as}\ \  \rho\rightarrow\infty  
\ee
and is therefore normalizable on the whole line $(-\infty,+\infty)$. 
Thus in the black hole case there exists a zero energy  bound  state 
too. The function $\psi_{0}$ possesses $n$ nodes,  which  shows  the 
existence of $n$ negative energy bound states. 
From here we conclude that for each $(n,r_{h})$ EYM black hole
solution, $r_{h}\in (0,\infty)$, there exist precisely $n$
negative modes in the odd-parity spherical perturbation sector.
For the $n=1$ EYM black holes, the negative modes energy 
eigenvalues for the both parity types are presented numerically
in Tab.1.


\vspace{3 mm}
\begin{center}

Tab.1. The bound state energies for the $n=1$ EYM black holes.

\vspace{5 mm}

\begin{tabular}{|c|c|c|} \hline

$r_{h}$    & $\omega^{2}$ (even-parity) & $\omega^{2}$ (odd-parity)\\ \hline

0 (regular case) &  --0.05249    & --0.06190    \\ \hline

0.1              &  --0.05026    &  --0.06182   \\ \hline

0.5              &  --0.04021    &  --0.06000   \\ \hline

1                &  --0.02683    &  --0.04926   \\ \hline

5                &  --0.00212    &  --0.00298   \\ \hline

10               &  --0.00054    &  --0.00075   \\ \hline
\end{tabular}

\end{center}

\vspace{3 mm}

Finally, we consider the case of the charged $SU(2)\times U(1)$  EYM 
black holes \cite{charge} which possess the additional  $U(1)$  dyon 
type field. In this case, the only new thing  appears  --  an  extra 
term $q^{2}/2r^{2}$ in the right hand side  of  Eq.(\ref{8}),  where 
$q^{2}=q^{2}_{e}+q^{2}_{m}$, $q_{e}$  and  $q_{m}$  being  the  dyon 
electric and magnetic charges; the odd-parity perturbation equations 
remain intact. The background solutions are qualitatively  the  same 
as in the $q=0$ case, but the presence of  the  charge  entails  now 
that $r_{h}\geq q$. When $r_{h}>q$, all of  the  analysis  performed 
above for the $q=0$ case remains valid. The only  difference  arises 
in the extreme limit, $r_{h}=q$, when $N$ has zero of  second  order 
at the horizon, and $f^{2}(r_{h})=1$. After all, the change  of  the 
boundary condition at the horizon results in the following change in 
the asymptotic behaviour for $\psi_{0}$ given by (\ref{36}): 
\be
\psi_{0}\sim\frac{1}{\rho^{2}}\ \ {\rm as}\ \ \rho\rightarrow -\infty, 
\ \ \ \ \ \  \ \
\psi_{0}\sim\frac{1}{\rho^{2}}\ \ {\rm as}\ \  \rho\rightarrow\infty.  
\ee
One can understand this as follows:  for  extreme  black  holes  the 
geometry in the limit  $\rho\rightarrow  -\infty$  is  known  to  be 
conformally  flat.  The  dynamics  of  the   conformally   invariant 
Yang-Mills field then must be the same as that in the asymptotically 
flat region, $\rho\rightarrow \infty$ (remind  that  the  odd-parity 
modes involve perturbations of the  Yang-Mills  field  alone).  Thus 
$\psi_{0}$ is normalizable in the case as  well,  and  we  therefore 
conclude that all charged $SU(2)\times U(1)$ black holes,  including 
extreme solutions, also possess $n$  odd-parity  spherical  negative 
modes.

\section*{Acknowledgments}
I  would  like  to  thank  Professor  N.Straumann
and also O.Brodbeck for discussions.

This  work
was supported by the Swiss National Science Foundation.

\appendix
\renewcommand{\thesection}{Appendix \Alph{section}}
\renewcommand{\theequation}{\Alph{section}\arabic{equation}}


\section{Variation of the energy functional}
\setcounter{equation}{0}

In this Appendix second variation of the ADM mass functional
presented in the main text is explicitly derived. The 
analysis is carried out for the regular BK solution,
the generalization on the black hole case is
straightforward. 

Consider the barrier height functional defined by Eq.(\ref{s49}):
\be
\varepsilon[f(r)]   =    \int_{0}^{\infty}(f'^{2}+\frac{(f^{2}-1)^{2}}
{2r^{2}})exp(-2\int_{r}^{\infty}f'^{2}\frac{dr}{r})dr. \label{B1}
\ee
Let  $f(r)$  be  a  smooth  function  satisfying   the
following conditions
\be
f'(r)=O(r)\ \ \ {\rm as}\ r\rightarrow 0;\ \ \ \
f'(r)=O(1/r^{2})\ \ \ {\rm as}\ r\rightarrow\infty.       \label{B2}
\ee
{\it Define} for convenience two new functions
\be
\sigma(r)=exp\{-2\int_{r}^{\infty}f'^{2}\frac{dr}{r}\} ,\ \ \
m(r)=\frac{1}{\sigma(r)}\int_{0}^{r}(f'^{2} +
\frac{(f^{2}-1)^{2}}{r^{2}})\sigma dr,                  \label{B3}
\ee
they satisfy the following boundary conditions
\be
\sigma(\infty)=1,\ \ \ \sigma(0)\neq 0;\ \ \
 m(\infty)<\infty,\ \ \ m(r)=O(r^{3})\ \ {\rm as}\ r\rightarrow 0,
                                                        \label{B4}
\ee
and also, by definition, the following equations
\be
\sigma'=2\frac{f'^{2}}{r}\sigma,\ \ \ \
 (m\sigma)'=(f'^{2}+\frac{(f^{2}-1)^{2}}{2r^{2}})\sigma. \label{B5}
\ee
Consider small variation
\be f(r)\ \rightarrow\ f(r)+\varphi (r),                \label{B6}
\ee
where
\be \varphi (0)=\varphi(\infty )=0.                     \label{B7}
\ee
To preserve the boundary conditions, the variation must also satisfy
\be \varphi'(r)=O(r)\ \ {\rm as}\ r\rightarrow 0,
\ \ \ \varphi'(r)=O(1/r^{2})\ \ {\rm as}\ r\rightarrow\infty.
                                                          \label{B8}
\ee
Putting (\ref{B6}) into (\ref{B1}) and expanding the result over
$\varphi$,  one obtains
\be
\varepsilon[f+\varphi]=\varepsilon[f]+\delta\varepsilon+\delta^{2}
\varepsilon + \ldots ,                                      \label{B9}
\ee
where the first variation is
\be  \delta\varepsilon       =       2\int_{0}^{\infty}\{f'\f'\sigma
+
\frac{f(f^{2}-1)}{r^{2}}\sigma\f     -2I(f'^{2}+\frac{(f^{2}-1)^{2}}
{2r^{2}})\sigma\} dr,                                   \label{B10}
\ee
and the second variation is
\be \delta^{2}\varepsilon = \int_{0}^{\infty}
\{\sigma\f^{2}+\sigma\frac{3f^{2}-1}{r^{2}}\f^{2} -
8I(f'\f'+\frac{f(f^{2}-1)}{r^{2}}\f)\sigma + $$
$$+\ (8I^{2}-2J)(f'^{2}+\frac{(f^{2}-1)^{2}}{2r^{2}})\sigma\}dr,
                                                        \label{B11}
\ee
dots  in  (\ref{B9})   denote   higher   order   terms,   and    the
following  new functions have been introduced:
\be I(r)=\int_{r}^{\infty}f'\f'\frac{dr}{r},\ \
J(r)=\int_{r}^{\infty}\f'^{2}\frac{dr}{r}.              \label{B12}
\ee
The boundary conditions (\ref{B2}),(\ref{B8}) imply that
\be I(0)<\infty,\ J(0)<\infty,\ I(\infty)=J(\infty)=0.  \label{B13}
\ee
Consider  the  first  variation   (\ref{B10}).   Using   (\ref{B5}),
   represent
$\delta\varepsilon$ as follows
\be
\delta\varepsilon =2\int_{0}^{\infty}\{f'\f'\sigma+
\frac{f(f^{2}-1)}{r^{2}}\f\sigma - 2I(m\sigma)'\}dr.   \label{B14}
\ee
Integrating by parts one has
\be
\left. \delta\varepsilon=(2f'\sigma\f-4Im\sigma)\right|_{0}^{\infty}+
2\int_{0}^{\infty}\{-(f'\sigma)'\f     +     \sigma\frac{f(f^{2}-1)}
{r^{2}}\f +2m\sigma I'\}dr.                            \label{B15}
\ee
Note, that the boundary terms in  this  expression  vanish.  Finding
$I'$ from (\ref{B12})  and  integrating  by  parts  once  more,  one
finally arrives at
\be
\delta\varepsilon=2\int_{0}^{\infty}\{
-((1-\frac{2m}{r})\sigma
f')'+\frac{f(f^{2}-1)}{r^{2}}\sigma\}\f  dr.\label{B16}
\ee
One  can  see  that  the
vanishing  of  the first variation implies the Yang-Mills equation
\be 
((1-\frac{2m}{r})\sigma f')'=\sigma\frac{f(f^{2}-1)}{r^{2}}.
                                                       \label{B17}
\ee

Assume now that the first variation vanishes and consider  the
second variation. Using (\ref{B5}) one obtains
$$
\delta^{2}\varepsilon=\int_{0}^{\infty}\{\sigma\f'^{
2}+ \sigma\frac{3f^{2}-1}{r^{2}}\f^{2}  -   8I(f'\f'
+  \frac{f(f^{2}-1)}
{r^{2}}\f)\sigma + $$
\be +\ 8I^{2}(m\sigma)'-2J(m\sigma)'\}dr.            \label{B18}
 \ee
Integrating the fourth term in the integrand  by  parts,  and  using
Eq.(\ref{B12}), one obtains
\be
\left.    8\int_{0}^{\infty}I^{2}(m\sigma)'dr    =     8I^{2}m\sigma
\right|_{0}^{\infty}   -    16\int_{0}^{\infty}II'm\sigma    dr    =
16\int_{0}^{\infty}Im\sigma f'\f'\frac{dr}{r} ,      \label{B19}
\ee
where the boundary terms vanish. Combining this result with the
third term in Eq.(\ref{B18}), one has
\be -8\int_{0}^{\infty}I\{ (1-\frac{2m}{r})\sigma        f'\f'    +
\frac{f(f^{2}-1)}{r^{2}}\sigma\f\} dr.               \label{B20}
\ee
Using (\ref{B17}),(\ref{B12}),(\ref{B5}), represent this expression
as follows
$$ -8\int_{0}^{\infty}I((1-\frac{2m}{r})\sigma f'\f)'dr =
\left.  -\  8I(1-\frac{2m}{r})\sigma     f'\f\right|_{0}^{\infty}  +
$$
$$
+\ 8\int_{0}^{\infty}I'(1-\frac{2m}{r})\sigma f'\f dr =
-8\int_{0}^{\infty}(1-\frac{2m}{r})\sigma              f'^{2}\f\f'
\frac{dr}{r}=$$
\be =-4\int_{0}^{\infty}(1-\frac{2m}{r})\sigma'\f\f' dr,\label{B21}
 \ee
where the boundary terms vanish. The fifth term in (\ref{B18})
yields
$$\left.  -2\int_{0}^{\infty}J(m\sigma)'dr  =  -2Jm\sigma\right|_{0}
^{\infty}+  2\int_{0}^{\infty}J'm\sigma  dr  =   -2\int_{0}^{\infty}
m\sigma\f'{2}\frac{dr}{r}= $$
\be = \left. -2m\sigma\frac{\f}{r}\f'\right|_{0}^{\infty} +
2\int_{0}^{\infty}(m\sigma\frac{\f'}{r})'\f dr  = \int_{0}^{\infty}
(\frac{2m}{r}\sigma\f')'\f dr.                      \label{B22}
\ee
The first term in (\ref{B18}) is
\be
\iii \sigma\f'^{2}dr=-\iii(\sigma\f')'dr,            \label{B23}
\ee
where  the  boundary  terms  are  zero.  Consider also the following
expression
\be 0=\iii((1-\frac{2m}{r})\sigma'\f^{2})'dr.        \label{B24}
\ee
Adding  the  equations    (\ref{B21})-(\ref{B24}) and
introducing  also  the tortoise coordinate, $\rho$,
\be \frac{dr}{d\rho}= \sigma(1-\frac{2m}{r}),     \label{B25}
\ee
one finally arrives at
\be \delta^{2}\varepsilon = \iii\f(-\frac{d^{2}}{d\rho^{2}} +V)\f\
d\rho,                                            \label{B26}
\ee
where
\be V=\sigma(1-\frac{2m}{r})\{ 2(\sigma'(1-\frac{2m}{r}))'       +
\frac{3f^{2}-1}{r^{2}}\sigma\},                        \label{B27}
\ee
which  agrees  with the potential in 
Eq.(\ref{12:1})  provided  that   Eqs.(\ref{B5}),
(\ref{B17}) are  taken  into account.

\end{document}